\documentclass[aps,prl,twocolumn,superscriptaddress]{revtex4}
\usepackage{graphicx}
\usepackage{amsmath}

\begin{document}


\title{Simulation of cohesive head-on collisions 
of thermally activated nanoclusters}


\author{Hiroto Kuninaka}
\email[E-mail: ]{kuninaka@phys.chuo-u.ac.jp}
\affiliation{Department of Physics, 
Chuo University, Bunkyo-ku, Tokyo, Japan, 112-8551}
\author{Hisao Hayakawa}
\affiliation{Yukawa Institute for Theoretical Physics, 
Kyoto University, Sakyo-ku, Kyoto, Japan, 606-8502
 }

\date{\today}

\begin{abstract}
 Impact phenomena of nanoclusters subject to thermal fluctuations 
 are numerically investigated. 
 From the molecular dynamics simulation for colliding two identical 
 clusters,  it is found that  the restitution coefficient 
 for head-on collisions has a peak at a colliding speed 
 due to the competition between the cohesive interaction and the repulsive interaction of colliding clusters. 
 Some aspects of the collisions can be understood by the theory by Brilliantov {\it et al.} 
(Phys. Rev. E {\bf 76}, 051302 (2007)), but 
many new aspects are found from the simulation.
In particular, we find that there are some anomalous rebounds in which 
the restitution coefficient is larger than unity.
 The phase diagrams of rebound processes 
 against impact speed and the cohesive parameter can be understood by a simple phenomenology. 
\end{abstract}


\maketitle
\section{Introduction}
Inelastic collisions are the process that a part of initial 
macroscopic energy of colliding bodies is distributed into 
the microscopic degrees of freedom. 
This irreversible process of head-on collisions 
may be characterized by the restitution coefficient 
which is the ratio of the normal rebound speed to the normal impact speed. 
Although it was generally believed that the restitution coefficient  is a material constant, 
modern experiments and simulations have revealed that the restitution coefficient  
decreases with the increase of impact velocity.\cite{vincent,raman,goldsmith,stronge} 
For example, in the case of collisions between icy particles, we can easily find the monotonic 
decrease of the restitution coefficient against impact velocity without any flat region.\cite{bridges} 
The dependence of the restitution coefficient on the low impact velocity is theoretically 
treated by the quasistatic theory\cite{kuwabara,morgado,brilliantov96,schwager,ramirez}. 
In Ref.\cite{kuwabara}, Kuwabara and Kono performed impact experiments by the use of 
a pendulum of various materials to validate their theoretical prediction.
On the other hand,  the dependence of the restitution coefficient on the high impact velocity 
is treated by the dimensional analysis based on plastic collisions.\cite{johnson}  
From the dimensional analysis, the relationship between the restitution coefficient $e$ and 
the impact velocity $V$ becomes $e \propto V^{-1/4}$, 
which coincides with experimental results 
by the use of a steel ball and blocks of various materials such as hard bronze and brass.
\cite{vincent,goldsmith,johnson}
We also recognize that the restitution coefficient can be less than unity 
for head-on collisions without any introduction of explicit dissipation, 
because the macroscopic inelasticities originate in the 
transfer of the energy from the translational mode to the 
internal modes such as vibrations.\cite{morgado,gerl,ces}

Although it is believed that  the restitution coefficient for head-on collisions 
is smaller than unity, the restitution coefficient can be larger than unity 
in oblique collisions.\cite{louge,kuninaka_prl,calsamiglia}
For example, Louge and Adams observed such an anomalous impact in which the restitution coefficient is larger than unity in oblique collisions of a hard 
aluminum oxide sphere onto a thick elastoplastic polycarbonate plate in which
 the restitution coefficient increases monotonically with the increase of 
the magnitude of the tangent of the angle of incidence.\cite{louge} 
They explained that this phenomena can be attributed to the change in rebound 
angle resulting from the local deformation of the contact area between the 
sphere and the plate, which causes the increase in the normal component of 
the rebound velocity against the collision plane. 
The present authors performed a two-dimensional impact 
simulation with an elastic disc and an elastic wall consisted of 
nonlinear spring network to reproduce the anomalous impacts.
They also explained the mechanism to appear large restitution coefficient based on 
 a simple phenomenology  
by taking into account the local surface deformation. \cite{kuninaka_prl}


The static interaction between macroscopic granular particles is characterized 
by the Hertzian theory\cite{hertz,landau} of the elastic repulsive force
as well as the dissipative force which is 
proportional to relative speed of colliding two particles. 
The total force acting between granular particles in contact is assumed to be 
a combination of the elastic repulsive force and the dissipative force in 
the quasistatic theory, with which many aspects of the inelastic collisions for 
such granular particles can be understood.  
This theory can reproduce  the restitution coefficient as a function 
of the colliding speed observed in experiments and simulations.\cite{kuwabara,brilliantov96,kuninaka-hayakawa2006}

Although the repulsive interaction becomes dominant 
for collisions of large bodies, 
cohesive interactions  such as van der Waals force and electrostatic force
play important roles for small clusters of the nanoscale.\cite{surf,castellanos,tomas}
Recently, Brilliantov {\it et al.}  have developed the quasistatic 
theory for inelastic collisions to explain
the relation between the colliding speed and 
the restitution coefficient for cohesive collisions.\cite{brilliantov07} 
The result of 
an experimental result of  collisions of macroscopic particles 
with the cohesive interaction is consistent with the theory.\cite{louge2008}

For molecular dynamics simulations of small clusters, 
many  empirical potentials are  used to mimic 
the interaction between various atoms.\cite{rieth} 
Among them, most commonly used  one is  the Lennard-Jones potential:
\begin{equation}\label{LJ}
U(r_{ij})=4\epsilon\left\{\left(\frac{\sigma}{r_{ij}}\right)^{12}-
               \left(\frac{\sigma}{r_{ij}}\right)^{6}\right\}, 
\end{equation}
which well approximates the interaction 
between inert gas atoms such as argons.\cite{rieth,allen} 
Here, $r_{ij}$ is the distance between two atoms 
labeled by $i$ and $j$, respectively. 
$\epsilon$ and $\sigma$ are the energy constant and the characteristic diameter, 
respectively. In this potential, the second term on the right hand 
side represents the cohesive interaction which is originated 
from van-der Waals interaction.

Dynamics of nanoclusters 
are extensively investigated from both scientific and technological interests. 
There are a lot of studies on cluster-cluster and 
cluster-surface collisions based on the molecular dynamics simulation.\cite{kalweit,kalweit06,knopse,tomsic,harbich} 
We observe variety of rebound processes in such systems 
caused by the competition between  
the attractive interaction and the repulsive interaction 
of two colliding bodies. 
Binary collisions of identical clusters cause
coalescence,  scattering, and fragmentation  
depending on the cluster size and the impact energy.
\cite{kalweit,kalweit06} 
On the other hand, cluster-surface collisions induce
soft landing, embedding, 
and fragmentation.\cite{harbich}  The attractive interaction plays crucially important roles in such colliding processes.

However, the attractive interaction may be reduced in the case of some 
combinations of the two interacting objects and the relative 
configuration of colliding molecules~\cite{sakiyama}. 
Awasthi {\it et al.} carried out the molecular dynamics simulation 
for collisions of Lennard-Jones clusters onto surfaces to simulate 
the collision of a $\rm{Bi}$ cluster onto a 
$\rm{SiO_{2}}$ surface.\cite{awasthi}  
They introduced a cohesive parameter to characterize 
the magnitude of attraction 
and investigate the rebound behavior of the clusters. 
Similarly, recent papers have reported 
that surface-passivated Si nanoclusters exhibit elastic rebounds on 
Si surface due to the reduction of the attractive interaction between the surfaces.\cite{suri,hawa} 
These results suggest that nearly repulsive collisions  really exist
even in small systems. 

In the case of purely repulsive collisions between two identical 
nanoclusters, we have already reported that the relation 
between colliding speed and the restitution coefficient 
may be described by the quasi-static theory for inelastic impacts, though the restitution coefficient exceeds unity for small impact speed.\cite{ptps,apm2008}  
In addition, on the basis of the distribution function of macroscopic energy 
loss during collision, we have shown that our numerical results 
can be approximated by the fluctuation relation for inelastic impacts.\cite{ptps}


The aim of the present paper is to study statistical properties 
in binary head-on collisions of identical nanoclusters.  
In particular, we numerically investigate the effects of attractive 
interaction on the restitution coefficient in rebound processes. 
The organization of this paper is as follows. In the next section, 
we introduce our numerical model of colliding nanoclusters and the setup of 
our simulation. 
In Section III, we summarize the results of our simulation.
In Section IV, we mainly discuss the system size dependence of our results. 
In Section V, we summarize our results. 
Appendices A, B, and C treat the calculation of 
the surface tension, the technical calculation on the system size dependence of the restitution coefficient, and  stability of spherical shape 
of a elastic droplet, respectively.

\section{Model}
\begin{figure}[htbp]
\begin{center}
\includegraphics[width=.2\textwidth]{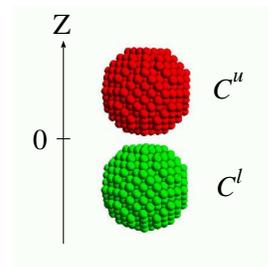}
\end{center}
\caption{(Color online) A typical situation of our simulation of two colliding clusters. 
Each of them contains 682 ``atoms'' which are bound by the Lennard-Jones potential.
}
\label{fig1}
\end{figure}

Let us introduce our numerical model. Our model consists of 
two identical clusters, each of which is spherically cut from 
a face-centered cubic (SC-FCC) lattice  of ``atoms''. 
We typically use 682 atoms systems which are 
$13$ layers SC-FCCs. The system size dependence will be discussed in Section IV. 
Here, we list the relation between the number of ``atoms'' and 
the number of layers in one cluster in Table~\ref{lay}. 
The clusters have facets due to the small number of ``atoms'' 
(Fig. ~\ref{fig1}). All the ``atoms'' in each cluster are bound 
together by the Lennard-Jones potential $U(r_{ij})$ in Eq.(\ref{LJ}).
When we regard the ``atom'' as argon, the values of the constants become 
$\epsilon=1.65\times10^{-21}\mathrm{J}$ and $\sigma=3.4$\AA, 
respectively.~\cite{rieth}
%
%
 \begin{table}
 \caption{\label{lay}Relation between numbers of layers and atoms.}
 \begin{ruledtabular}
 \begin{tabular}{cc}
 Number of Layers & Number of Atoms \\
\hline
3 &   12\\
5 &   42\\
7 &  135\\
9 &  236\\
11 &  433\\
13 &  682\\
15 & 1055\\
17 & 1466
 \end{tabular}
 \end{ruledtabular}
 \end{table}
%

Henceforth, we label the upper and the lower clusters as 
 $C^{u}$ and  $C^{l}$, respectively. 
We assume that the interactive potential  between the atom $k$ on the lower surface of $C^{u}$ 
and the atom  $l$ on the upper surface of $C^{l}$ is given by
\begin{equation}
\varphi(r_{kl})=4\epsilon\left\{\left(\frac{\sigma}{r_{kl}}\right)^{12}-
               c \left(\frac{\sigma}{r_{kl}}\right)^{6}\right\}, 
\end{equation}
where $r_{kl}$ is the distance between the surface atom $k$ and atom $l$. 
We introduce the cohesive parameter $c$  to characterize 
the attraction between the atoms of different clusters.\cite{awasthi} 

The procedure of our simulation is as follows. 
As the initial condition of simulation, 
the centers of mass of $C^{u}$  and $C^{l}$ are placed 
along the $z$-axis with the separation  $\sigma_{c}$ between 
the surfaces of $C^{u}$  and $C^{l}$. 
The initial velocities of the ``atoms'' in both $C^{u}$ and $C^{l}$ 
obey Maxwell-Boltzmann distribution with the initial temperature $T$. 
The initial temperature is set to be $T=0.02 \epsilon$ 
in our simulations.  
Sample average is taken over different sets of initial velocities 
governed by the Maxwell-Boltzmann velocity distribution for ``atoms''. 

\begin{figure}[htbp]
\begin{center}
\includegraphics[width=.45\textwidth]{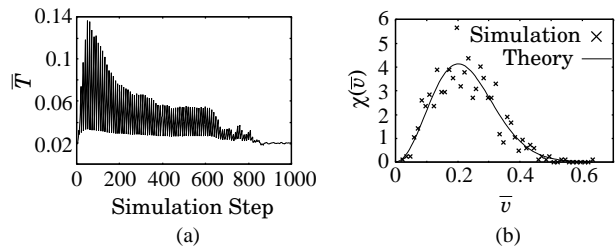}
\end{center}
\caption{
(a) Relaxation of kinetic temperature and (b) distribution of speed of atoms 
after equilibration process. $\bar T$ and $\bar v$ are
temperature and speed of ``atom'' scaled by units $\epsilon$ 
and $\sqrt{\epsilon/m}$, respectively. 
}
\label{fig1-2}
\end{figure}

To equilibrate the clusters, we adopt the velocity scaling 
method~\cite{woodcock,nose} and perform $2000$ steps simulation 
for the relaxation to a local equilibrium state. Here let us check the equilibration of 
the total energy in the initial relaxation process. 
Figure \ref{fig1-2}(a) is the time evolution of the kinetic temperature of $C^{u}$, where $\bar T$ denotes the scaled temperature by the unit $\epsilon$. 
This figure shows the convergence of temperature to the desired temperature $T=0.02\epsilon$. 
On the other hand, Fig.\ref{fig1-2}(b) is the probability density distribution of speed of  ``atoms" in $C^{u}$ 
when the equilibration process is over, where $\bar v$ denotes the scaled 
velocity for ``atom" by the unit $\sqrt{\epsilon/m}$.  
The solid curve in Fig.\ref{fig1-2}(b) shows 
the probability density distribution of speed $v_{i}$ of ``atoms" indexed by $i$ in equilibrium, 
\begin{equation}
\chi(v_i)=4 \pi \left(\frac{m}{2\pi k T}\right)^{3/2} v_{i}^{2}\exp\left(-\frac{m}{2kT}v_{i}^{2}\right),
\end{equation}
with $T=0.02\epsilon$. This agreement  shows that the upper cluster $C^{u}$ is equilibrated 
during the equilibration process.

After the equilibration, 
we give translational velocities to $C^{u}$ and $C^{l}$ 
to make them collide against each other. 
The relative speed of impact 
ranges from $V=0.02 \sqrt{\epsilon/m}$ to $V=0.6 \sqrt{\epsilon/m}$.
Here, the characteristic speed is 
the thermal velocity for one ``atom'' 
 $\sqrt{T/m}$, where $m$ is the mass of each ``atom''. 
This situation might correspond to the sputtering process or collisions of interstellar dusts or atmospheric dusts.
Although it is not easy to control the velocity of colliding clusters  in real nanoscale experiments,  
the effects of thermal fluctuation to the center of mass of each cluster
might be negligible if clusters are flying in vacuum. 

Numerical integration of the equation of motion for each atom 
is carried out by the second order symplectic integrator with 
the time step $dt=1.0 \times 10^{-2} \sigma/\sqrt{\epsilon/m}$. 
To reduce computational costs, we introduce the cut-off length $\sigma_{c}=2.5 \sigma$ 
of the Lennard-Jones interaction, which sometimes affects 
the energy conservation of a system although the Hamiltonian of the system is conserved. 
 We have checked that the rate of energy conservation, $|E(t)-E_{0}|/|E_{0}|$, 
is kept within $10^{-5}$ with the cutoff length $\sigma_{c}=2.5 \sigma$, 
where $E_{0}$ is the initial energy of the system and $E(t)$ is the 
energy at time $t$. In general, the value between $3\sigma \le \sigma_{c} \le 4\sigma$ 
is used 
for the energy conservation about $|E(t)-E_{0}|/|E_{0}| \sim 10^{-5}$.

We let the angle around $z-$axis, $\theta_{z}$, be $\theta_{z}=0$ 
when the two clusters are located mirror-symmetrically 
with respect to $z=0$. 
In most of our simulation, we set $\theta_{z}$ at $\theta_{z}=0$ 
as the initial condition. 
The dependency on $\theta_{z}$ will be shown in the next section.

\section{Results of our simulation}

\subsection{Relation between impact speed and restitution coefficient}

\begin{figure}[htbp]
\begin{center}
\includegraphics[width=.4\textwidth]{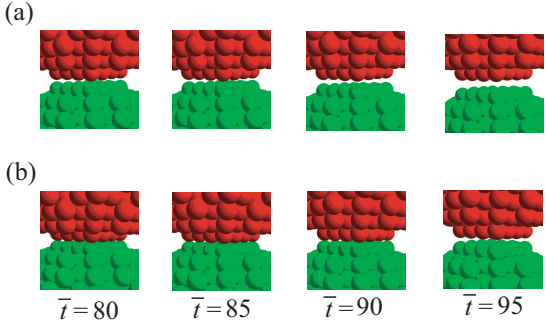}
\end{center}
\caption{
 (Color online) Sequential plot of collisions for (a) c=0.0 and (b) c=0.2 
at $\bar t=80$, $85$, $90$, and $95$, where $\bar t$ is time scaled 
by unit $\sigma/\sqrt{\epsilon/m}$. 
}
\label{KH-fig3}
\end{figure}

Figures \ref{KH-fig3} (a) and (b) display, respectively, the magnified sequential plots of colliding 
clusters for a purely repulsive collision and a cohesive collision 
when the initial temperature 
and the impact speed are $T=0.02\epsilon$ and $V=0.3\sqrt{\epsilon/m}$. 
From Fig. \ref{KH-fig3},  we confirm that the contact duration for the cohesive collision is longer than that 
of the repulsive collision.\cite{awasthi} 
During the restitution, we also observe the elongation of the clusters 
along the $z$-axis in cohesive collisions,  
while we can not observe such a phenomenon in repulsive collisions. 
In both cases, the rotation of clusters is slightly excited after a collision.

\begin{figure}[htbp]
\begin{center}
\includegraphics[width=.35\textwidth]{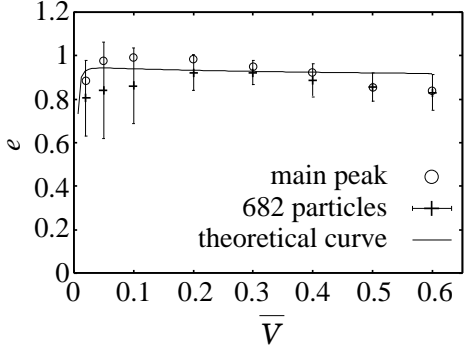}
\end{center}
\caption{
Relationship between colliding speed and restitution coefficient 
for $c=0.2$. $\bar V$ is relative colliding speed scaled 
by unit $\sqrt{\epsilon/m}$.
}
\label{KH-fig5}
\end{figure}
We firstly investigate the relation between the colliding speed 
and the restitution coefficient for a weak attractive case ($c=0.2$).  
The cross points in Fig.~\ref{KH-fig5} show the relationship between 
the relative speed of impact scaled by the unit $\sqrt{\epsilon/m}$, 
$\bar V \equiv V/ \sqrt{\epsilon/m}$,  
and the restitution coefficient $e$. 
Sample average is taken over 100 different initial conditions for each speed and the error 
bar shows the standard deviation. 
From Fig.~\ref{KH-fig5}  we find that the restitution coefficient has a peak around 
the colliding speed $V = 0.2 \sqrt{\epsilon/m}$, which is attributed 
to the reduction of $e$ caused by the attractive force in the lower impact speed. 
This tendency can also be observed  in cohesive collisions of 
macroscopic bodies.~\cite{brilliantov07,louge2008}

The solid line in Fig.~\ref{KH-fig5}  represents the  theoretical prediction of cohesive 
collisions between viscoelastic spheres.\cite{brilliantov07} 
Here, we briefly summarize the theory of cohesive collisions in Ref.\cite{brilliantov07}. 
Let us consider a head-on 
collision between elastic spheres of radii $R_{1}$ and $R_{2}$, each of which has mass of 
$M_{1}$ and $M_{2}$, respectively.
The basic idea of their theory is to solve the time evolution equation of the deformation 
$\xi(t)$ of the colliding spheres:
\begin{eqnarray}\label{org}
\mu \ddot \xi(t) + F(\xi(t)) =0, \\\notag
\xi(0)=0,
\hspace{3mm}
\dot \xi(0)=V,
\end{eqnarray}
where $\mu$ is 
the reduced mass $\mu=(1/M_1+1/M_2)^{-1}$. 
$\xi(t)$ is described as the function of the radius of 
contact area $a$ as
\begin{equation}
\xi(a)=\frac{a^2}{R_{eff}}-\displaystyle\sqrt{\frac{8\pi\gamma D a}{3}} 
\hspace{3mm} 
\text{with}
\hspace{3mm} 
 R_{eff}=\left(\frac{1}{R_1}+\frac{1}{R_2}\right)^{-1}, 
\end{equation}
so that Eqs.(\ref{org}) are rewritten as 
\begin{eqnarray}\label{brilliantov_ode} 
 \mu \ddot a +\mu \frac{\xi^{''}(a)}{\xi^{'}(a)} \dot a^{2}
    + \frac{F(a)}{\xi^{'}(a)} = 0, \\
        a(0)=a_{init},
        \hspace{3mm} 
\dot a(0)=V \left(\frac{d\xi}{da}|_{a_{init}}\right)^{-1},
\end{eqnarray}
where the prime denotes the differentiation with respect to $a$. 
We adopt $a_{init}=(8 \pi D \gamma R_{eff}^{2}/3)^{1/3}$ 
which is the contact radius of the bottom plane of the upper cluster with 
$\gamma \simeq 0.104 \epsilon/\sigma^{2}$ estimated from 
the calculation of the attractive interaction between two clusters 
(see Appendix~\ref{appA} ). We also estimate $D$ as 
$D=3.28 \times 10^{-3} \sigma^{3}/\epsilon$ 
from Young's modulus $Y=454\epsilon \sigma^{-3}$ and Poisson's ratio 
$\nu=7.74 \times 10^{-2}$ 
which are obtained from another simulation.~\cite{ptps} 

They assume that the force $F(a)$ between cohesive spheres 
comprises three kinds of forces: elastic force $F_{H}(a)$ characterized by Hertzian contact theory\cite{hertz,landau}, 
dissipative force $F_{dis}(a)$\cite{kuwabara}, 
and cohesive Boussinesq force $F_{B}(a)$ derived 
from JKR theory.\cite{jkr} 
Thus, the total force can be expressed by 
\begin{equation}
F(a)=F_{H}(a)-F_{B}(a)+F_{dis}(a).
\end{equation}
Here, the sum of the elastic force and the Boussinesq force is given by
\begin{equation}
F_{H}(a)-F_{B}(a)=\frac{a^2}{R_{eff}}-\displaystyle\sqrt{\frac{6\pi\gamma}{D}}a^{3/2}
\end{equation}
with the surface tension $\gamma$
and $D=(3/4)\{(1-\nu_{1}^{2})/2Y_1+(1-\nu_{2}^{2})/2Y_{2}\}$ with Poisson's ratio $\nu_i$ and Young's modulus $Y_i$ 
for the cluster $i=1,2$.
Following the idea in Ref.\cite{brilliantov07}, we assume that the dissipative force is given by
\begin{equation}
F_{dis}(a)=A \dot{a}\frac{\partial}{\partial a}(F_{H}(a)-F_{B}(a)),
\end{equation}
where $A$ is a fitting parameter.

We solve Eq.(\ref{brilliantov_ode}) with the initial speed $V$ ranging from 
$0.01(\epsilon/m)^{1/2}$ to $0.6(\epsilon/m)^{1/2}$ 
by the use of the fourth order Runge-Kutta method to 
obtain the rebound speed which is the speed 
when the contact radius $a$ becomes less than 
$a_{sep} \equiv \left(3 \pi D \gamma R^{2}_{eff}/2\right)^{1/3}$.\cite{brilliantov07}
From the rebound speed for each impact speed, we obtain the relationship between 
 the restitution coefficient and the impact speed.  
In Fig.~\ref{KH-fig5}, 
we use $A=0.1 \sigma\sqrt{m/\epsilon}$ to draw the theoretical curve. 
In $V < 0.2 \sqrt{\epsilon/m}$, 
the discrepancy between our numerical results (cross points) and the theoretical result 
is large, which may be attributed to the rotational rebounds of 
clusters after collisions.  On the other hand, 
the theoretical curve reproduces
the results of simulation which excludes rotation of clusters (open circles)  as will be explained later.

\begin{figure}[htbp]
\begin{center}
\includegraphics[width=.35\textwidth]{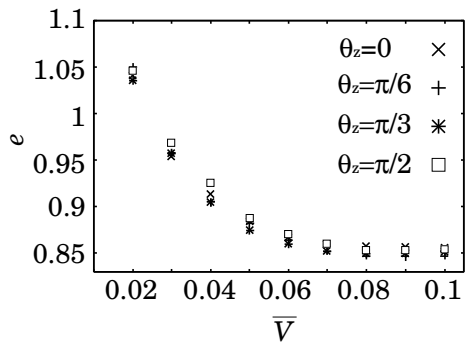}
\end{center}
\caption{
Relation between relative colliding speed and restitution coefficient 
for different initial angles for $c=0.0$ and $T=0.02\epsilon$.
}
\label{kakudo-izon}
\end{figure}

Here, let us briefly comment on the dependence of 
the relative angle $\theta_{z}$ on the numerical results. 
We have checked  $\theta_z$ dependence of the restitution for purely repulsive collisions, 
{\it i.e.} $c=0.0$ at $T=0.02\epsilon$. 
Figure \ref{kakudo-izon} shows the relationship between 
impact speeds and restitution coefficients for 
$\theta_{z}=0, \pi/6,  \pi/3,\hspace{1mm}\text{and}\hspace{1mm} \pi/2$. 
This figure indicates that the relation between the impact speed and 
the restitution coefficient is not largely affected by the initial 
orientation, 
although the orientation around other axes may affect the relation. 
Thus, we will analyse only the results obtained with the fixed initial 
orientation $\theta_{z}=0$.

\subsection{Distribution of restitution coefficient}

\begin{figure}[t]
\begin{center}
 \begin{minipage}{0.47\textwidth}
  \includegraphics[width=0.72\textwidth]{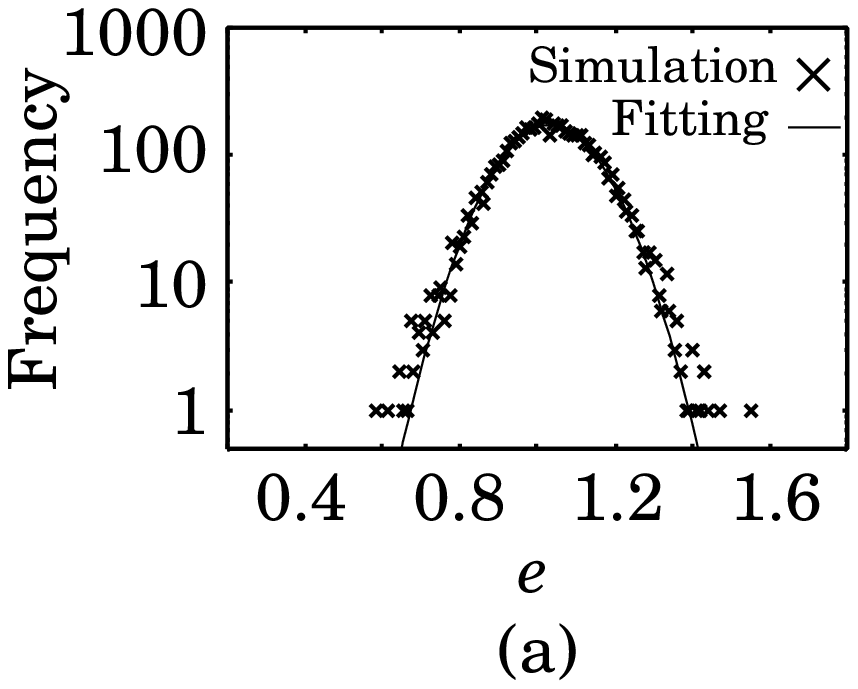}
 \end{minipage}
\hspace*{3mm}
 \begin{minipage}{0.47\textwidth} 
  \includegraphics[width=0.7\textwidth]{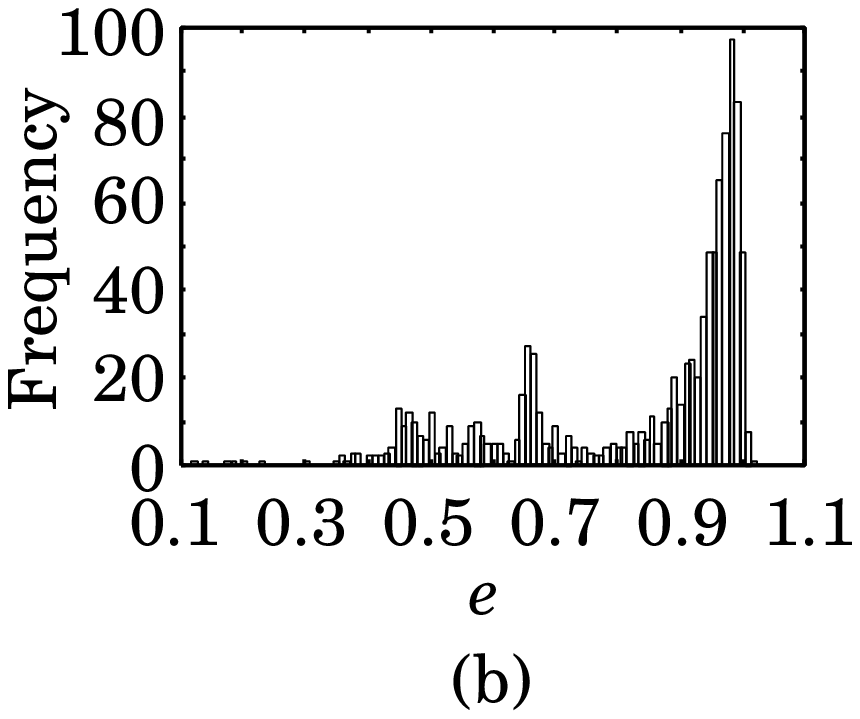}
 \end{minipage}
\caption{Histograms of restitution coefficients for (a) $c=0.0$, 
 $V=0.02\sqrt{\epsilon/m}$, and (b) $c=0.2$, $V=0.1\sqrt{\epsilon/m}$. 
The solid line in (a) is the Gaussian fitting of the data.
}
\label{hist-a0_2-v0_1}
\end{center}
\end{figure}

In this subsection, we investigate the frequency distributions of restitution 
coefficients for purely repulsive collisions and cohesive collisions, respectively. 
Figure \ref{hist-a0_2-v0_1}(a) shows the histogram of 
the restitution coefficient for purely repulsive collisions ($c=0.0$). 
To obtain this result, we take $5000$ samples at the 
fixed impact speed $V=0.02 \sqrt{\epsilon/m}$. 
From Fig. \ref{hist-a0_2-v0_1}(a),  the frequency distribution can be 
approximated by the Gaussian (solid line) for purely repulsive collisions.

On the other hand,  Fig. \ref{hist-a0_2-v0_1}(b) shows 
the frequency distribution of the restitution coefficient 
for cohesive collisions ($c=0.2$). 
To obtain this result, we take $995$ samples at the fixed impact speed 
$V=0.1 \sqrt{\epsilon/m}$. 
In Fig. \ref{hist-a0_2-v0_1}(b), 
we find the existence of the two peaks around $e=0.448$ and $e=0.656$, 
respectively, except for the main peak around $e=0.982$.
From the check of simulation movies,
the collisions around these small peaks 
are produced by  rotations  after the collisions, 
while the most of bounces are not associated with rotations 
in the vicinity of the main peak around $e=0.982$. 
It is reasonable that the excitation of macroscopic rotation lowers 
the restitution coefficient.

Here, let us make another comparison of our simulation result 
with the theoretical curve drawn in Fig.\ref{KH-fig5}. 
The open circles in Fig.\ref{KH-fig5} are the mean values  
obtained by the data around  the main peak for each impact speed to remove the effects of rotational bounces.
It is obvious that  the theory has a better fitting curve of the data when we remove  rotational bounces.

We shall comment on the fitting function of the main peak.  
Figure \ref{hist-a0_2-v0_1}(b) shows that the distribution around the main peak 
has an asymmetric profile, so that the Gaussian function may fail to 
fit the tail parts of the main peak. 
\begin{figure}[htbp]
\begin{center}
\includegraphics[width=0.36\textwidth]{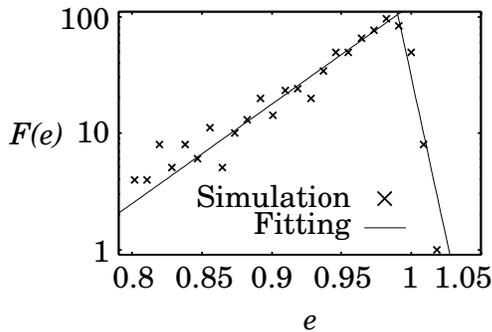}
\caption{Semi-log plot of the main peak in Fig.\ref{hist-a0_2-v0_1}(b). 
The solid lines are the double exponential functions to fit the data. 
}
\label{semi}
\end{center}
\end{figure} 
Figure \ref{semi} shows the semi-log plot of the simulation data around the main peak, where 
$F(e)$ is the frequency of $e$. 
The reasonable fitting curves are represented by the solid lines, 
where $\ln F(e)=(19.6 \pm 1.9) e +(-14.7 \pm 1.8)$ for $e<0.982$ and 
$\ln F(e)=(-127 \pm 28) e +(130 \pm 28)$ for $e>0.982$, respectively. 
Thus, the distribution of restitution coefficients can be approximated 
by a combination of the double-exponential functions 
when the rotation is not excited after impacts. 
A similar tendency has also been observed  
in a recent experiment and a recent simulation of macroscopic collisions.
\cite{poeschel2008}

\subsection{Phase diagram of restitution coefficient}

As discussed in the previous subsection, some samples of the restitution coefficient exceeds unity even for cohesive collisions.
We can guess that most of colliding clusters coalesce when we use the collisional model with $c=1$. Thus, it is important to know what
process actually occurs after a collision when the impact speed or the cohesive parameter $c$ is given. 
In this subsection, we investigate the emergence probability of  four modes of the collisions (i) coalescence, 
(ii) bouncing,
(iii) normal collision with $e<1$, and
(iv) anomalous collision with $e>1$. 
The coalescence (i) and the bouncing (ii) can take place only when the attractive interaction 
between the colliding clusters exists. 
Indeed, the bouncing occurs  as the result of trapping by the potential well, if the rebound speed is not large enough.
\cite{awasthi06}

\begin{figure}[htbp]
\begin{center}
\includegraphics[width=.5\textwidth]{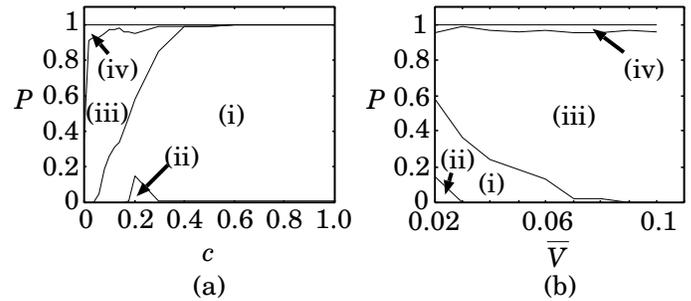}
\end{center}
\caption{
Probability diagrams classified by four collision modes 
with fixing (a) $V=0.02\sqrt{\epsilon/m}$ and 
(b) $c=0.2$, respectively.
}
\label{phase_diag}
\end{figure}

Figure \ref{phase_diag} (a) shows the phase diagram which is obtained 
under the fixed colliding speed $V=0.02\sqrt{\epsilon/m}$, where
$P$ represents the probability to observe each mode. 
This phase diagram shows that the regions for the modes (iii) and (iv) decrease 
with the increase of $c$. In the strong attractive case,  $c>0.6$, we cannot observe rebound 
modes (ii), (iii), and (iv).
Here, we  find that the anomalous impact can be observed 
for cohesive collisions with $c<0.4$.

We also  categorize  collisions into four modes 
as a function of the impact speed under the fixed cohesive parameter $c=0.2$ 
(Fig.\ref{phase_diag} (b)). Here, we find that the probability to emerge 
the modes (i) and (ii) decreases with the increase of the impact speed. 
In addition, 
the anomalous impact can be observed within the range 
of impact speed $0.02 \le V/(\epsilon/m)^{1/2} \le 0.1$. 
It is interesting that Fig.\ref{phase_diag}(b) for $V<0.04\sqrt{\epsilon/m}$ is almost the mirror symmetric 
one of Fig.\ref{phase_diag}(a) for $c<0.2$, 
which suggests that the cohesive parameter plays a role of the impact speed.

Here let us reproduce the results of our simulation qualitatively by a phenomenology.
Purely repulsive collisions, 
as we expect from Fig.~\ref{hist-a0_2-v0_1}(a), 
the probability density distribution of rebound speed $V^{'}$ 
can be approximated by a Gaussian function 
\begin{equation} \label{gauss}
p(V^{'})=\frac{1}{\sqrt{\pi \alpha}} 
\exp\left[-\frac{(V^{'}-V_{m})^2}{\alpha}\right].
\end{equation}
Thus, for given impact speed $V$, we can use Eq.(\ref{gauss})
where
$V_{m}$ and $\alpha$ in Eq.(~\ref{gauss}) are  fitting parameters. 
Then, we calculate the probability  
to exceed the escape speed $V^{*}$~\cite{barger} 
from the attractive potential field  
$\Phi(r) \equiv -4\epsilon c (r/\sigma)^{-6}$ under the given $V'$. 
Here the escape speed $V^{*}$ of a rebounded cluster 
may be given by
\begin{equation}\label{escape}
 V^{*}/\sqrt{\epsilon/m}=\sqrt{2\Phi(x^{*}/\sigma)/\sigma\mu},
\end{equation}
where $x^{*}=(2/c)^{1/6}\sigma$ 
at which the potential takes the minimum value. 
For example, the escape speed becomes $V^{*}=0.015 (\epsilon/m)^{1/2}$ 
in the case of $c=0.2$. 

Thus, from integrating the probability densities 
of $V^{'}$, 
the probabilities to observe modes (i), (iii), and (iv) 
are respectively given by 
\begin{widetext}
\begin{eqnarray}
P^{(\rm{i})} &=& \int_{-\infty}^{V^{*}} p(V^{'}) dV^{'}= 
 \frac{1}{2}\left\{1-{\rm erf} 
\left(\frac{V_m-V^{*}}{\sqrt{\alpha}}\right)\right\},\label{p1}\\
P^{(\rm{iii})} &=& \int_{V^{*}}^{V} p(V^{'}) dV^{'}= 
 \frac{1}{2}\left\{1-{\rm erf} \left(\frac{V_m-V}{\sqrt{\alpha}}\right)\right\}
-P^{\rm{(i)}},\label{p2}\\
P^{\rm{(iv)}} &=& \int_{V}^{\infty} p(V^{'}) dV^{'} = 1-P^{\rm{(i)}}-P^{\rm{(iii)}}\label{p3},
\end{eqnarray}
\end{widetext}
where ${\rm erf}(x)$ is the error function 
${\rm erf}(x)\equiv \int_{-\infty}^{x} \exp(-t^{2}) dt$. 
Here we ignore the distinction between the mode (i) and the mode (ii), because the most of bouncing clusters eventually
coalesce after some numbers of collisions.

\begin{figure}[t]
\begin{center}
\includegraphics[width=.5\textwidth]{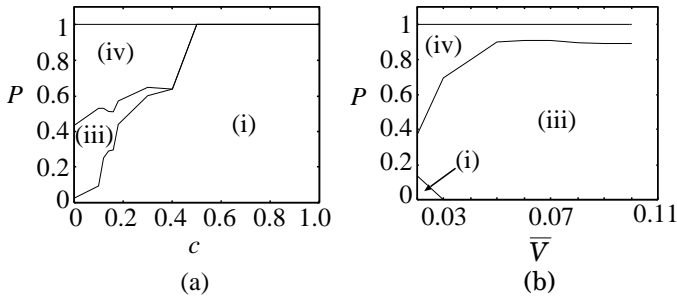}
\end{center}
\caption{
Probability diagrams from our theoretical argument 
for (a) $V=0.02\sqrt{\epsilon/m}$ and (b) $c=0.2$.
}
\label{phase_theor}
\end{figure}

Figures \ref{phase_theor} (a) and (b) show the probability diagrams 
obtained from Eqs. (\ref{p1}), (\ref{p2}), and (\ref{p3}). 
To draw Fig.\ref{phase_theor}(b), 
we adopt $V^{*}=0.018 \sqrt{\epsilon/m}$ which 
is slightly larger than the calculated value  $V^{*}=0.015 \sqrt{\epsilon/m}$  by Eq.(\ref{escape}) for $c=0.2$. 
Our phenomenology qualitatively reproduces the diagrams obtained by the simulation as 
in Figs.\ref{phase_diag} (a) and (b), although there are some quantitative differences between the simulation and the phenomenology.
Indeed,  the probability to appear the mode (iv) in the phenomenology 
decreases with the increase of $V \sqrt{\epsilon/m}$ 
as in Fig.\ref{phase_theor} (b), but this tendency cannot be observed in the simulation in Fig.\ref{phase_diag} (b).

\section{Discussion}
\begin{figure}[t]
\begin{center}
\includegraphics[width=.35\textwidth]{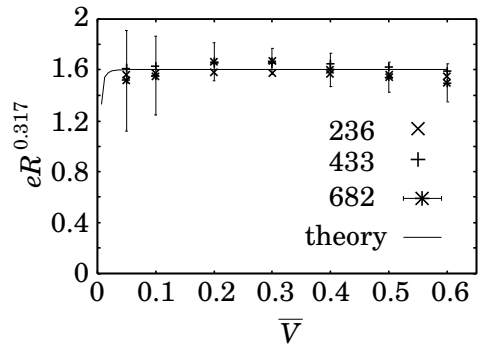}
\end{center}
\caption{
Relation between colliding speed and restitution coefficient 
for clusters $\rm{C}_{236}$, $\rm{C}_{433}$, and $\rm{C}_{682}$. 
The restitution coefficients are scaled by $R^{0.317}$, where $R$ is 
the radius of each cluster.
}
\label{size2}
\end{figure}

%
%
Let us discuss our results. We, mainly, discuss how 
the restitution coefficient depends on the cluster size in this section. 
Figure \ref{size2} shows the relationship 
between the relative colliding speed of clusters and the restitution coefficient with different sizes of 236 atoms ($\rm{C}_{236}$), 
433 atoms ($\rm{C}_{433}$), and 682 atoms ($\rm{C}_{682}$), respectively, 
where we use the data obtained by the fixed parameters $c=0.2$ 
and $T=0.02 \epsilon$.  
As can be seen in Fig. \ref{size2}, the restitution coefficients satisfies the scaling in which 
$e (R/\sigma)^{0.317}$ is a universal function of the impact speed,
where $R$  is the radius of each cluster. 
To obtain the scaling exponent, 
we first calculate the standard deviation for each 
rebound speed under the fixed value of the exponent. Next, we search the value of 
the exponent such that the maximum value of the standard deviations has a minimum value. 
To draw the solid curve in Fig.\ref{size2}, 
we solve eq.(\ref{brilliantov_ode}) with the fitting parameter 
$A=0.1 \sigma\sqrt{m /\epsilon}$ for $\rm{C}_{682}$ with the aid of its radius $R_{eff}=3.23\sigma$.
This is interesting finding from our simulation which has not been predicted 
by the quasistatic theory of cohesive collisions for macroscopic bodies.\cite{brilliantov07}

From a simple phenomenology, we can understand that the restitution coefficient depends on the radius.
However, the phenomenology predicts that $e(R/\sigma)^{1/2}$ satisfies a scaling relation (see Appendix \ref{appB}). 
The discrepancy between the phenomenology and the numerical observation indicates that
our over-simplified theory is insufficient. We will need a more sophisticated theory
to explain the exponent.

\begin{figure}[bht]
\begin{center}
\includegraphics[width=.47\textwidth]{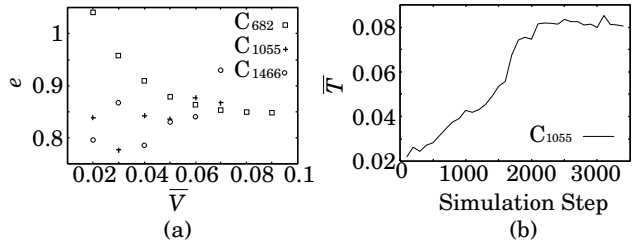}
\end{center}
\caption{
(a) Relation between impact speed and restitution coefficient 
in cases of $\rm{C}_{682}$, $\rm{C}_{1055}$, and $\rm{C}_{1466}$. 
(b) Time evolution of internal temperature of cluster $\rm{C}_{1055}$ 
in its free flight after initial equilibration to $T=0.02\epsilon$. 
}
\label{large}
\end{figure}

%
%
We also simulate collisions between larger clusters than $\rm{C}_{682}$ 
by the use of $\rm{C}_{1055}$ and $\rm{C}_{1466}$. 
Figure \ref{large} (a) is the relationship between the impact speed 
and the restitution coefficients 
in the case of $c=0.0$ under the initial temperature $T=0.02\epsilon$. 
The squares, plus points, and circles show the averaged data 
of $\rm{C}_{682}$, $\rm{C}_{1055}$, $\rm{C}_{1466}$, respectively.  
We take $10$ samples for both $\rm{C}_{1055}$ and $\rm{C}_{1466}$ 
while $100$ samples for $\rm{C}_{682}$.
Here we do not find any systematic relationship between the impact speed 
and the restitution coefficients in the cases of $\rm{C}_{1055}$ and 
$\rm{C}_{1466}$. 
This can be attributed to the surface instability of 
the clusters arising from the weak attraction between ``atoms''. 

It is known that the instability of the spherical shape and the plastic deformation in a cluster
cause the increase of the internal temperature of the cluster.\cite{awasthi,suri} 
We numerically performed free flights of cluster by the use of $\rm{C}_{1055}$ 
to check the time evolution of the internal temperature of the cluster. 
Figure \ref{large}(b) shows the time evolution of the temperature 
inside the cluster $\rm{C}_{1055}$ after giving the translational speed 
$V=0.07 \sqrt{\epsilon/m}$ and the initial temperature $T=0.02\epsilon$.  
Here we find the temperature increase during the free flight  up to 
around $T=0.08\epsilon$. 
Thus, we conclude that the maximum number of ``atom'' to reproduce 
the theory of cohesive collision is $682$ in our system.  

We try to estimate the critical radius theoretically on the 
basis of the argument of capillary instability of elastic droplets 
(see Appendix \ref{appC}),\cite{landau2}  but 
 our over-simplified theory predicts that any elastic surface 
of spheres are unstable under the gravity.
We should note that this calculation is based on theory of elasticity 
with zero shear modulus (Poisson's ratio is equal to -1).
The calculation suggests that (i) we may not use theory of elasticity 
or (ii) zero shear modulus is unrealistic.  
We will, at least, need to discuss the capillary instability 
under the full set of elastic equations.  
From these arguments, we regard $\rm{C}_{682}$ as the maximum size 
to reproduce the quasi-static theory of cohesive collisions 
in our modelling. 

%
%
Although our simulation mimics impact phenomena 
of small systems subject to large thermal fluctuations, 
we should address that our model with small $c$ may not be adequate 
for the description of many of realistic collisions of nanoclusters, 
where the cohesive interaction between 
clusters often prohibits the rebound in the low-speed impact.
Namely, the corresponding value of the cohesive parameter is large in many of actual situations. 
However, nanoscale impacts can be realized experimentally by the using 
the surface coated nanoclusters. 
For example, it has been demonstrated that hydrogen coated Si
nanoparticles exhibit the weak attraction by H atoms on the surface.\cite{suri,hawa}
We believe that our model captures the essence of such a system. 
For realistic simulations, we may need to carry out another simulation of 
the collision of H-passivated Si clusters by introducing suitable 
empirical potentials. 
As an additional remark, we should indicate that 
it is difficult to control the colliding speed and
the initial rotation of the cluster in actual situations 
because the macroscopic motion of one cluster is also affected 
by thermal fluctuations.

\section{Conclusion}

In conclusion, we have performed molecular dynamics simulations 
to investigate the behaviors of colliding clusters and 
the relationship between the restitution coefficient and 
the impact speed. 
The results of our simulations have revealed that some aspects of the relationship 
can be understood by the quasi-static theory for cohesive collisions.\cite{brilliantov07} 
In addition, we have drawn the phase diagram of the restitution
coefficient in terms of the impact speed and the cohesive parameter 
and explained them by a simple phenomenology. 
To clarify the mechanism of the emergence of the anomalous impact, 
it may need further investigation about the internal state of clusters 
during collision such as stress and modal analyses. 

\acknowledgments
We would like to thank N.~V.~Brilliantov, T.~P\"oschel, 
M. Otsuki, T. Mitsudo, and K. Saitoh for their valuable comments. 
We would also like to thank M.~Y.~Louge 
for introducing us his recent experimental results. 
Parts of numerical computation in this work were carried out in computers of 
Yukawa Institute for Theoretical Physics, Kyoto University. 
This work was supported by the Grant-in-Aid for the Global COE Program 
"The Next Generation of Physics, Spun from Universality and Emergence" 
from the Ministry of Education, Culture, Sports, 
Science and Technology (MEXT) of Japan. This study is partially 
supported by the Grant-in-Aid of 
Ministry of Education, Science and Culture, Japan (Grant No. 18540371).

\vspace{3mm}
\appendix
\section{Calculation of surface tension}\label{appA}
In this appendix we explain how we calculate 
the surface tension $\gamma$ used 
to draw the theoretical curve in Fig.~\ref{KH-fig5}.\cite{surf} 
Let us assume that two identical clusters are in plane-to-plane contact 
with each other. When those clusters are located by the separation 
$d$, the surface energy per unit area $W$  is given by\cite{surf} 
\begin{equation}
W \simeq \frac{B}{12 \pi d_{0}^{2}}\left(1-\frac{d_{0}^{2}}{d^{2}}\right),
\end{equation}
where $B$ is the Hamaker constant 
$B \equiv 4 \pi^{2} \epsilon c \sigma^{6} \rho^{2}$ 
with the cohesive parameter $c$ and 
the number density $\rho$ of each cluster. 
In our model, we use the number density becomes $\rho=0.4\sigma^{-3}$, and 
$d_{0} \simeq 0.4\sigma$. 

The surface tension $\gamma$ is equal to the energy per unit area to separate 
the two contacting plane to infinity. Thus, we obtain $\gamma$ as 
\begin{equation}
\gamma =\frac{B}{24 \pi d_{0}^2}
\simeq 0.0261 \times \frac{4\epsilon}{\sigma^{2}}
=0.1044 \frac{\epsilon}{\sigma^2}.
\end{equation}

\section{Dependence of $e$ on $R$}\label{appB}
Let us derive a scaling relation between the restitution coefficient and 
the radius of cluster. 
Our assumption is that (i) the energy dissipation during a collision is originated from the sum of 
of the viscous force and the Boussinesq force, 
(ii) energy dissipation from the Boussinesq force is approximately given by 
the work during the detachment process of two coalesced clusters.  

\begin{figure}[htbp]
\begin{center}
\includegraphics[width=.2\textwidth]{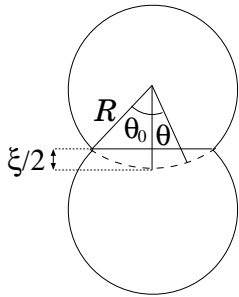}
\end{center}
\caption{
Schematic figure of contacting identical spheres.
}
\label{ct}
\end{figure}

Let us first estimate the energy dissipation caused by the surface tension.
A pair of colliding clusters is partially coalesced as shown in
Figure \ref{ct}, where we assume that the deformation of two spheres are negligible, and contacted state can 
be characterized by a simple cut of the deformed region.
Let $\theta$ be the angle around the center of the upper sphere 
ranging from $-\theta_{0}$ to $\theta_{0}$ under the assumption a small 
$\theta_{0}$. 
From this figure, the surface area of a cut hemi-sphere is approximately given by 
$\pi R^{2} \theta^{2}_{0}=\pi R \xi$ with the deformation 
$\xi \simeq \theta_{0}^{2} R$. Thus, the work needed to pull off two spheres is  
\begin{equation}\label{se}
W_{\gamma} \simeq 2 \pi \gamma R \xi_{max} \sim \left(\frac{\rho_0}{Y}\right)^{2/5} 
R^{2} V^{4/5},
\end{equation}
where we use the estimation $\xi_{max} \propto(\rho_0 R^3/Y \sqrt{R})^{2/5} V^{4/5} \sim
(\rho_0/Y)^{2/5}RV^{4/5}$ on the basis of the theory of elasticity, where $\rho_0$ is the density.\cite{landau}

On the other hand, the energy dissipation of repulsive spheres is given by\cite{kuwabara}
\begin{equation}\label{e-loss}
E_{loss}^{dis}\propto \rho_0^{3/5}\tau Y^{2/5} R^{2}V^{11/5},
\end{equation} 
where $\tau$ is time scale of the dissipation.
From the combination of two terms in eqs. (\ref{se}) and (\ref{e-loss}), we obtain the expression
of the total energy loss during a collision
\begin{equation}\label{total-loss}
E_{loss}=E_{loss}^{dis}-W_{\gamma}
\sim R^2(c_1 V^{11/5}-c_2 V^{4/5}),
\end{equation}
where $c_1\simeq \rho_0^{3/5}\tau Y^{2/5}$ and $c_2\simeq (\rho_0/Y)^{2/5}$.
Since Eq.(\ref{total-loss}) should be balanced with 
$\rho_0 R^3V^{2} (1-e^{2})$, we obtain 
\begin{equation}
R(1-e^{2}) \sim \frac{1}{\rho_0}(c_1 V^{1/5}-c_2V^{-6/5}).
\end{equation}
Thus, our phenomenology suggests that $R^{1/2}e$ is independent of the radius of the colliding spheres. 

\section{Instability of an elastic droplet}\label{appC}
In this appendix we investigate the instability of the surface 
profile of clusters on the assumption that the internal modes of 
the cluster are expressed by those of an isotropic elastic sphere. 
When the shear stress can be ignored, the stress tensor $\sigma_{ij}$ 
can be written as 
\begin{equation}
\sigma_{ij}=K \nabla \cdot {\bf u} \delta_{ij},
\end{equation}
where $K$ is the bulk modulus, $\delta_{ij}$ is Kronecker delta, and 
${\bf u}$ is the strain. 
Thus, the equation of motion in the bulk becomes 
\begin{equation}\label{eq3}
\rho \Ddot {\bf u}=\nabla(K \nabla \cdot {\bf u})=-\nabla p,
\end{equation}
where $\rho$ is the density and $p \equiv -K \nabla \cdot {\bf u}$ is 
the effective pressure. Thus, the problem can be mapped onto a problem of perfect fluid.
Thus, the dispersion relation is linearized equation $R(\theta,\phi,t)=R_0+\zeta(\theta,\phi,t)$ can be
written as
\begin{equation}
    \omega^{2}_{l}=\frac{\gamma l(l-1)(l+2)}{\rho R^{3}_{0}},
\end{equation}
as in the case of  a liquid droplet\cite{landau2}, where $l$ is the index of Legendre polynomial.

When we introduce the gravity in this perfect fluid model,
the scalar potential $\Phi$ defined 
by ${\bf v}=\nabla \Phi$ satisfies
\begin{equation}\label{eq7}
    \partial_{t} \Phi + P + \frac{1}{2}v^{2} +gz =f(t),
\end{equation}
where where $\partial_{t}$ is the time derivative, $P=\int dp/\rho(p)$, and 
$f(t)$ is an arbitrary function of time,
 $g$ and $z$ are the gravitational acceleration and 
the relative vertical position from the center of mass of the sphere. 
Choosing $f(t)$ satisfying $f(t)=p_0+\gamma \left(\frac{1}{R_1}+\frac{1}{R_2}\right)$ with the surface tension $\gamma$, curvatures $R_1$ and $R_2$,
Eq. (\ref{eq7}) can be rewritten as
\begin{equation}\label{eq8}
\rho \ddot \Phi
= \frac{\gamma}{R^{2}_{0}} \left\{2 \frac{\partial \Phi}{\partial r}
-\Lambda(\theta,\phi)\frac{\partial \Phi}{\partial r}
\right\}+\rho g \cos\theta \frac{\partial \Phi}{\partial r}. 
\end{equation}
where
$\Lambda(\theta, \phi) = -\left\{\frac{1}{\sin\theta}
\frac{\partial}{\partial \theta}
\left(\sin\theta \frac{\partial}{\partial \theta}\right)
+\frac{1}{\sin^{2}\theta}\frac{\partial^{2}}{\partial \phi^{2}}
\right\}$. 
To derive Eq. (\ref{eq8}) we have used $\dot \Psi = \Phi$, 
${\bf v} = \dot{\bf u}=\nabla \Psi$, and $\dot\zeta=v_r=\partial \Phi/\partial r$ at $r=R_0$.
 
By using the expansion $\Phi$ in terms of $r^l$ and the
spherical harmonic function $Y_{l,m}(\theta,\phi)$, we may
obtain
\begin{widetext}
\begin{equation}\label{3.5}
\omega_{l,m}^2=\frac{\gamma}{\rho R_0^3}l(l-1)(l+2)
-\frac{l g}{R_0}\left\{
\sqrt{\frac{(l-m)(l+m)}{(2l-1)(2l+1)}}
+
\sqrt{\frac{(l-m+1)(l+m+1)}{(2l+1)(2l+3)}}\right\},
\end{equation}
\end{widetext}
where we have used
the formula
\begin{multline}\label{3.4}
\cos\theta Y_{lm}=\displaystyle\sqrt{\frac{(l-m+1)(l+m+1)}{(2l+1)(2l+3)}}Y_{l+1,m}\\
+\displaystyle\sqrt{\frac{(l-m)(l+m)}{(2l-1)(2l+1)}}Y_{l-1,m}.
\end{multline}
Therefore, $\omega_{n,l,m}$ becomes complex, if the radius exceeds the critical radius
\begin{widetext}
\begin{equation}\label{3.6}
R_{0,c}^{(l,m)}
=\left[
\frac{\gamma(l-1)(l+2)}
{\rho g 
\left\{
\displaystyle
\sqrt{\frac{(l-m)(l+m)}{(2l-1)(2l+1)}}+
\displaystyle
\sqrt{\frac{(l-m+1)(l+m+1)}{(2l+1)(2l+3)}}
\right\}
}
\right]^{1/2}.
\end{equation}
\end{widetext}
Equation \eqref{3.6} implies that the mode with $l=1$ is always unstable for the perturbation.
Thus, we conclude that an accelerated elastic sphere is unstable, which is similar to the instability
of a raindrop of the perfect fluid because the neutral mode $l=1$ does not have any recovering force.

\end{document}